\documentclass[conference]{IEEEtran}
\IEEEoverridecommandlockouts
\usepackage{cite}
\usepackage{amsmath,amssymb,amsfonts}
\usepackage{algorithmic}
\usepackage{graphicx}
\usepackage{textcomp}

\newcommand{\refig}[1]        {Fig.~\ref{#1}}
\newcommand{\resec}[1]        {Sec.~\ref{#1}}
\DeclareMathOperator*{\argmin}{argmin}

\def\BibTeX{{\rm B\kern-.05em{\sc i\kern-.025em b}\kern-.08em
    T\kern-.1667em\lower.7ex\hbox{E}\kern-.125emX}}
\begin{document}

\title{Measurements and System Identification for the Characterization of Smooth Muscle Cell Dynamics \\
\thanks{This work is financially supported by Public Private Partnership Allowance made available by Health-Holland, Top-Sector Life Sciences and Health, to the Association of Collaborating Health Foundations (SGF) to stimulate public-private partnerships, and by ZonMW, grant LSHM21078-SGF “CELLSYSTEMICS”.}
}

\author{\IEEEauthorblockN{Dilan \"{O}zt\"{u}rk}
\IEEEauthorblockA{\textit{Control Systems Group} \\
\textit{Eindhoven University of Technology}\\
Eindhoven, The Netherlands \\
d.ozturk.sener@tue.nl}
\and
\IEEEauthorblockN{Pepijn Saraber}
\IEEEauthorblockA{\textit{Cardiovascular Research Institute Maastricht (CARIM)}\\ 
\textit{Maastricht University}\\
Maastricht, The Netherlands \\
pepijn.saraber@maastrichtuniversity.nl}
\and
\IEEEauthorblockN{Kevin Bielawski}
\IEEEauthorblockA{\textit{Optics11Life} \\
Amsterdam, The Netherlands \\
kevin.bielawski@optics11life.com}
\and
\IEEEauthorblockN{Alessandro Giudici}
\IEEEauthorblockA{\textit{CARIM}\\ 
\textit{Maastricht University}\\
Maastricht, The Netherlands \\
a.giudici@maastrichtuniversity.nl}
\and
\IEEEauthorblockN{Leon Schurgers}
\IEEEauthorblockA{\textit{CARIM}\\ 
\textit{Maastricht University}\\
Maastricht, The Netherlands \\
l.schurgers@maastrichtuniversity.nl}
\and
\IEEEauthorblockN{Koen D. Reesink}
\IEEEauthorblockA{\textit{CARIM} \\
\textit{Maastricht University}\\
Maastricht, The Netherlands \\
k.reesink@maastrichtuniversity.nl}
\and
\IEEEauthorblockN{Maarten Schoukens}
\IEEEauthorblockA{\textit{Control Systems Group} \\
\textit{Eindhoven University of Technology}\\
Eindhoven, The Netherlands \\
m.schoukens@tue.nl}
}

\maketitle

\begin{abstract}

Biological tissue integrity is actively maintained by cells. It is essential to comprehend how cells accomplish this in order to stage tissue diseases. However, addressing the complexity of a cell's system of interrelated mechanisms poses a challenge. This necessitates a well-structured identification framework and an effective integration of measurements. Here we introduce the use of state-of-the-art frequency-domain system identification techniques combined with an indentation measurement platform to analyze the underlying mechanisms from the perspective of control system theory. The ultimate goal is to explore how mechanical and biological factors are related in induced Pluripotent Stem Cell-derived vascular smooth muscle cells. We study on the frequency-domain analysis for the investigation and characterization of cellular dynamics of smooth muscle cells from the measured data. The measurement model in this study exploits the availability of human tissue and samples, enabling fundamental investigations of vascular tissue disease.
This approach using human cell lines holds significant potential to decrease the necessity for animal-based safety and efficacy studies.
The focus of this review is to investigate the cellular dynamics underlying the myogenic response and to demonstrate the practicability of employing a nano-indentation measurement setup for the broadband frequency-domain characterization of induced Pluripotent Stem Cell-derived vascular smooth muscle cells. 

\end{abstract}

\begin{IEEEkeywords}
system identification, cell mechanobiology, smooth muscle cells, frequency domain analysis, signal processing 
\end{IEEEkeywords}

\section{Introduction}

Vascular diseases significantly impact global health and pose a financial burden on the healthcare system due to their vast range of manifestations. Conditions like hypertension and aortic aneurysms typically arise from a dysfunction of the blood vessel wall \cite{humphrey2014b}. The blood vessel wall is not passive at all, it contains cells responsible for creating the connective tissue, referred to as the extracellular matrix (ECM), during development and later for maintaining this tissue through periods of growth and adaptation. Elastic arteries consist of vascular smooth muscle cells (SMCs) and a specialized ECM that offers elasticity and strength \cite{cocciolone2018}. There exist three primary load-bearing components in the arterial wall: collagen, elastin and vascular smooth muscle \cite{reesink2019}. Elastin and collagen are structural ECM components, while vascular SMCs represent the most abundant cell type in the medial layer, which are essential in regulating the remodeling activities within the vessel wall \cite{jaminon2019}. Together with ECM's elastin, SMCs bear the pressure load and ensure that the vessel is compliant.   

SMCs located in the aorta and large arteries live in an extremely dynamic mechanical environment. Aortic SMCs react to mechanical stretching by contracting, termed as \textit{myogenic response}, which restores cell dimensions in the short term \cite{davis1992}. Yet, SMCs respond to changes in the long term by engaging in activities like ECM synthesis, reorientation and migration, during which they adjust their internal settings \cite{jaminon2019, lacolley2017}. This suggests a strong connection between the aortic tissue integrity \cite{humphrey2014b} and the myogenic response and contractile phenotype of SMCs \cite{petsophonsakul2019}.    
However, cell-based repair techniques have not yet addressed the complex system dynamics of SMCs, highlighting the necessity to determine which regulatory mechanisms act during different phases of a vessel's life cycle \cite{luo2020, wight2018}. 

Standard patient evaluation typically includes medical imaging techniques such as CT and MRI, genetic analysis and tissue histology \cite{milewicz2017}. However, the significance of SMC mechanobiology during onset and development of a disease is not routinely considered \cite{humphrey2008}. The long-term objective of this study is to determine how mechanical as well as biological factors interact in tissue-derived disease SMCs and same-patient induced Pluripotent Stem Cell (iPSC)-derived SMCs. 

The current literature on animal experiments stresses the relevance of mechanobiology at the cellular level \cite{humphrey2014a}. However, there remains a gap in effectively integrating these insights up to the tissue-, vessel- and patient-level. 
In the literature, there exist studies focused on the measurements using medical devices on cell biology and SMCs \cite{sato2018, yim2005}. 
Several studies including indentation measurements of SMC dynamics have also been conducted. For example, recently, atomic force microscopy nanoindentation has been used for the mapping of individual living human aortic SMCs \cite{petit2022}.  
In another study, mechanical properties of collagen fibrils and the effect on SMCs has been analyzed using nanoindentation and atomic force microscopy \cite{mcdaniel2007}.

Various methodologies for analyzing the identification of SMC dynamics have been developed. Previous studies suggest that black box identification techniques have been minimally utilized in the analysis of the mechanobiology of SMCs and multicellular populations. Conversely, numerous grey box mathematical models have been developed to characterize the mechanical behavior of SMCs \cite{kroon2010, tan2015}, but these models have mostly not been validated within a dynamic systems framework. An example study adopting a dynamical systems perspective on the mechanical characterization of SMCs has used the frequency response of muscle stiffness in single SMCs \cite{shue1999}. 
In terms of cell-substrate modeling, the proposed model for the biomechanical behavior of smooth muscle tissue is divided into two components, including active and passive components \cite{kroon2010}. 
In another study, the contractile units and extracellular components have been modeled, with only collagen fibers being considered, while other components were neglected \cite{tan2015}. The development of constitutive models for SMC contraction has generally based on well-established physiological knowledge on their contractile apparatus. 

In light of these studies, we focus on the development of a myogenic response identification framework for characterising the SMC dynamics. The myogenic response refers to the contraction of a blood vessel that is initiated by an increase in intravascular pressure \cite{meininger1992}. This response was firstly observed over a century ago by Bayliss and colleagues, who addressed changes in the volume of a dog's hindlimb after alteration of systemic blood pressure \cite{bayliss1902}. Later studies suggested that the circumferential tension in the vessel wall might be the controlling parameter for the myogenic response \cite{johnson1980}. The mechanisms underlying the myogenic response have been also investigated at the cellular level \cite{davis1999}. Many recent studies have examined the myogenic response based characteristics of vascular smooth muscle \cite{carlson2005, yang2003}. In the present work, we propose a novel method for characterizing SMC dynamics through the use of frequency-domain approximations derived from multisine signals, introducing an innovative approach. We study the system identification framework for the investigation and exploitation of cellular dynamics of SMCs. Our approach focuses on identifying the dynamics of myogenic response from a control system perspective. The contributions of this article include 
\begin{itemize}
\item [i.] the development of a measurement setup for the broadband frequency-domain characterization of the mechanical properties of SMCs using indentation measurements,
\item [ii.] the practical application of this measurement setup for analyzing the frequency-domain dynamics of SMCs. 
\end{itemize}

\section{Measurement Setup}
\label{sec:measurement setup}

Measurements were carried out using a Pavone nanoindenter (Optics11 Life, Amsterdam, NL) equipped with a special module developed to enable custom multisine signals to be generated as input to the system. The system performs the measurements by controlling the displacement of a piezo flexure that holds a cantilever probe. A probe stiffness of $0.021$ N/m and tip diameter of $19$ $\mu$m was used for the experiments. Cells were positioned under the sphere of the cantilever such that the tip was above the nucleus. The probe was positioned approximately $20$ $\mu$m above the cell, and a D-mode indentation \cite{mattei2015} was started with a holding period of $90$ seconds. After the piezo reached the target position, relaxation was allowed for approximately $2$ seconds, and the multisine signal was generated. The piezo then had a final holding period until the end of the indentation. Data on the cantilever bending (force) and the piezo displacement was acquired during the measurement. Indentation of the cells was assumed from the moment the cantilever bending reached a target threshold. 

\vspace{-0.2cm}
\begin{figure}[h]
	\centering
	\includegraphics[width=0.4\linewidth]{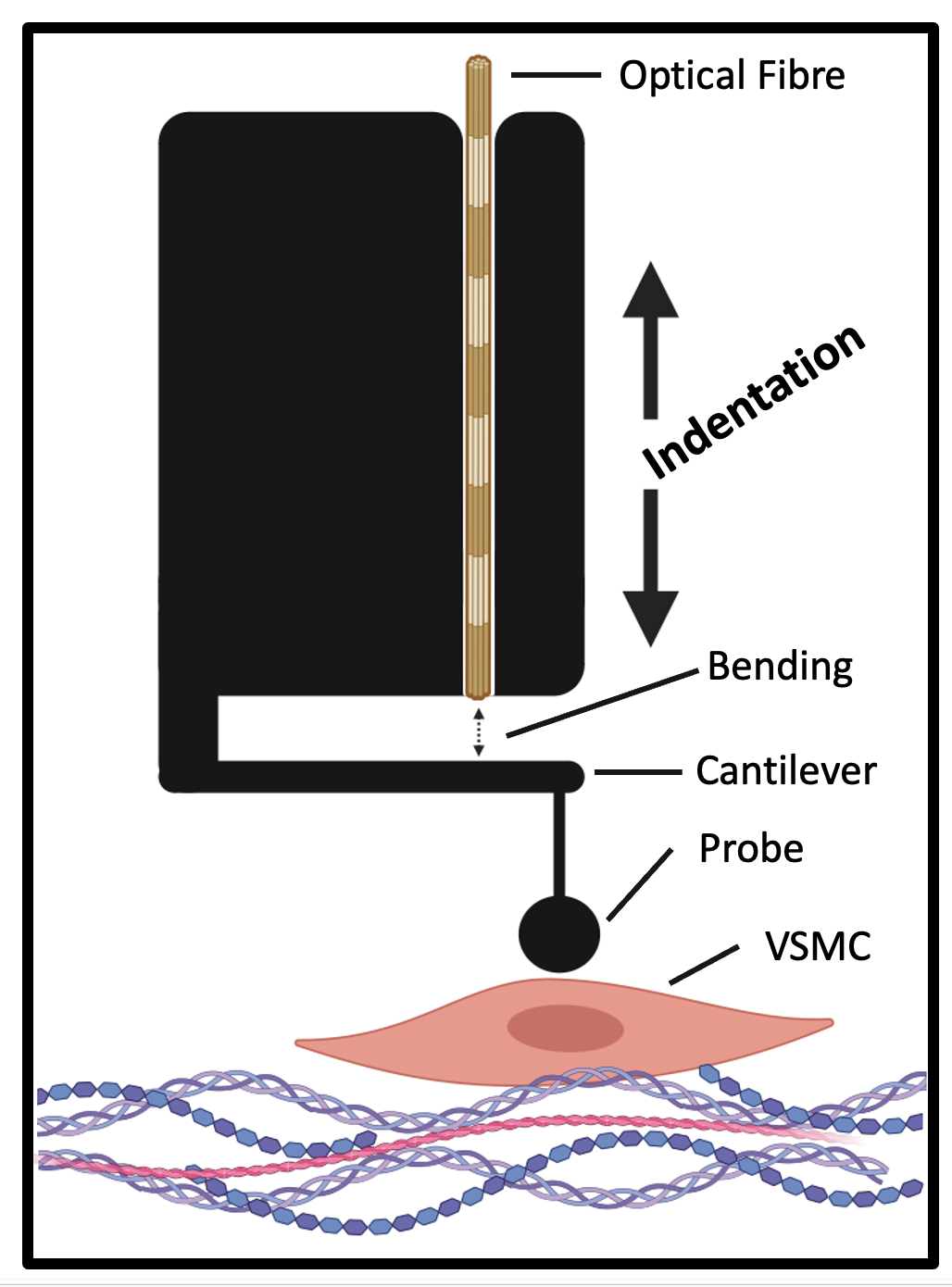}
	\caption{Nanoindentation on a cell, created with BioRender.}
	\label{fig:measurement}
\end{figure}

Other setups have also been used to measure the dynamic properties of cells with nanoindentation, typically using chirp signals or discrete frequencies \cite{efremov2020}, although some systems have also been used with a multi-frequency approach as well \cite{roca2006, takahashi2015}. Though this study investigated a small sample size, the multisine integration with the Pavone enables high throughput measuring the broadband dynamics of cells in future experiments. 
The nanoindentation setup on a cell is shown in \refig{fig:measurement}. 
\vspace{-0.5cm}
\begin{figure}[h]
	\centering
	\includegraphics[width=8cm]{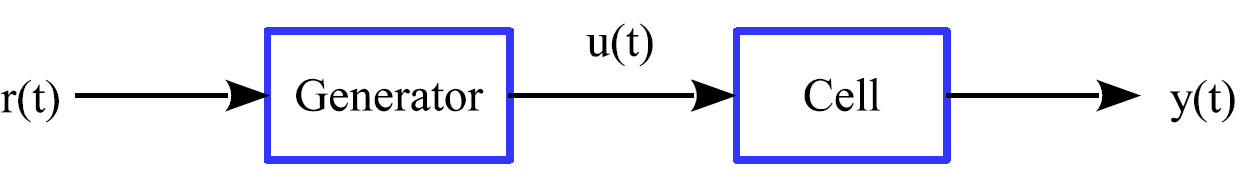}
	\caption{Measurement set up. r(t): reference multisine signal, u(t): load and y(t): indentation}
	\label{fig:system}
\end{figure}

\refig{fig:system} illustrates the measurement setup, detailing the inputs and outputs within the system. $r(t)$ is the reference multisine signal. The load ($u(t)$) and indentation ($y(t)$) are sampled at $f_s = 1000$ Hz and both include $N = 12800$ samples in one period. The reference signal $r(t)$ is generated at $1000/32$ Hz, using zero-order hold, and has a length of $N = 400$ samples in one period. This reference signal is given to the generator, which generates a displacement on the base of the cantilever, and the indentation on the cells is determined based on the difference between the movement of the base and the amount of deflection of the cantilever (or force applied to the sample).

\section{Identification of SMC Dynamics}

This section covers the identification of the dynamics at the level of the arterial wall. In particular, we focus on the quantification of the active contractile behavior of smooth muscle cells. Within the system identification field, linear time-invariant (LTI) systems are often characterized by measuring their frequency response function (FRF) \cite{schoukens2018}. When facing nonlinear dynamical systems, the FRF model is extended using the Best Linear Approximation (BLA) framework \cite{schoukens}, which allows for detecting and quantifying nonlinear distortions that are present in the measurements. 

\subsection{Identification Procedure}

The myogenic response is defined as the contraction of a blood vessel triggered by an increase in intravascular pressure \cite{meininger1992}. The primary motivation behind this study is to identify the dynamics of this active response, as we assume that such a response reinforces the contractile phenotype of the cell, thereby mitigating pathologies like aneurysms. We focus on identifying the indentation dynamics of single cells, including ECM and SMCs. Here the indentation replaces the pressure stimulus that cells sense \textit{in vivo}. The myogenic response is quantified by considering the load on the substrate with embedded SMCs as an input and the actively generated indentation as an output as shown in \refig{fig:system}. 

Using the measurement setup described in \resec{sec:measurement setup}, we examine the load and indentation data by utilizing a so-called random phase multisine signal over different frequency ranges. To facilitate the frequency-domain analysis of SMCs, the BLA of the model is investigated using the advanced FRF estimation approach implemented by the Local Polynomial Method (LPM) \cite{schoukens2018, schoukens2009}. This approach identifies the optimal linear time-invariant model (in least squares sense) that closely mimics the measured output of the nonlinear system. 

\subsection{Data Generation}

A random phase multisine signal is given by
\begin{equation}
	u(t) =  \sum_{k=1}^{N/2-1} A_k \cos(k \omega_0 t + \phi_k) \quad \text{with} \quad \mathbb{E} \{ e^{j\phi_k} \} = 0.
\end{equation}
In this paper, we utilized a linear frequency distribution in the specified range $[f_{min}, f_{max}]$, equal amplitudes $A_k$ over all excited frequencies, and a uniformly distributed $[0, 2\pi)$ phase $\phi_k$  \cite{schoukens}. $N$ is the number of samples in one period of the signal, and $\omega_0 = 2\pi f_0$, where $f_0 = f_s/N$ is the frequency resolution and $f_s$ is the sampling frequency considered during the signal generation.  

\begin{figure}[h]
	\centering
	\includegraphics[width=0.71\linewidth]{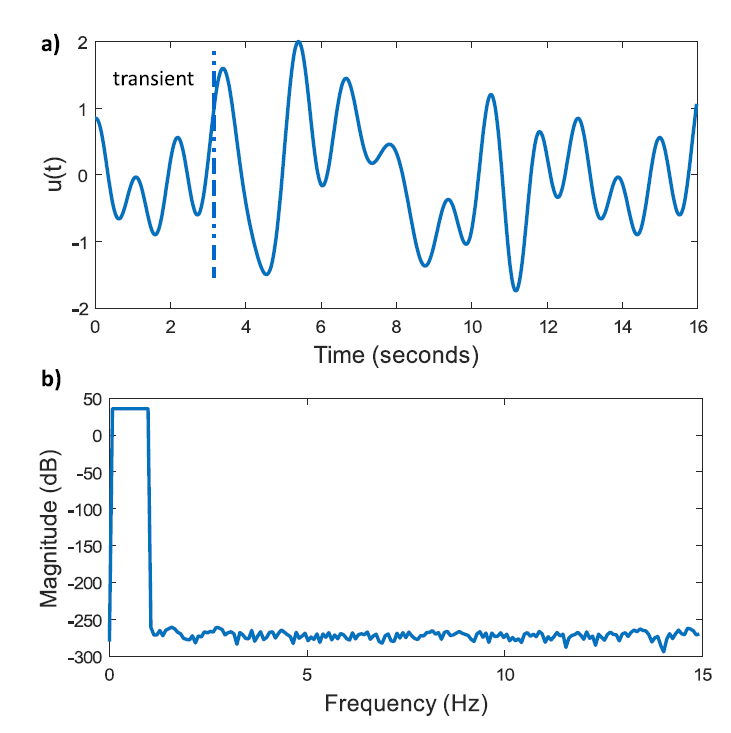}
	\caption{a) One period of the reference multisine signal. b) DFT spectrum of the reference multisine signal.}
	\label{fig:multisine}
\end{figure}

Multisine signals are employed as they offer the user full control over the amplitude spectrum of the excitation signal, and they allow the characterization of the cell dynamics at multiple frequencies simultaneously using a short experiment duration. This is of importance due to the degradation of the cells under test once an experiment starts. 
The random phase multisine reference signals in this work are generated with a sampling frequency of $f_s = 1000/32 = 31.25$ Hz and $N = 400$ samples in one period. The frequency resolution of the signal is $f_0 = f_s / N = 0.0781$ Hz. The considered excited frequency range is $[f_{min}, f_{max}] = [0.06 \, \text{Hz}, 1 \, \text{Hz}]$ (note that 1\,\text{Hz} is the physiological human heart rate at rest) and equal amplitudes $A_k = 0.02$ are considered. \refig{fig:multisine} (a) shows a single period of such a multisine signal in the time domain, while \refig{fig:multisine} (b) displays the discrete Fourier transform (DFT) spectrum of the reference multisine signal. Notably, at the onset of the signal, there are $100$ extra samples allocated from the last part of the signal to ensure a steady-state periodic nature of the obtained measurements, mitigating any leakage effects. Finally, multiple reference signals with different phase realizations of the multisine signal are used for the measurements on a single cell. This allows one to characterize the nonlinear behavior of the cells over the considered frequency range by using the BLA framework.

\subsection{FRF and BLA Estimation}

\begin{figure}[h]
	\centering
	\includegraphics[width=0.85\linewidth]{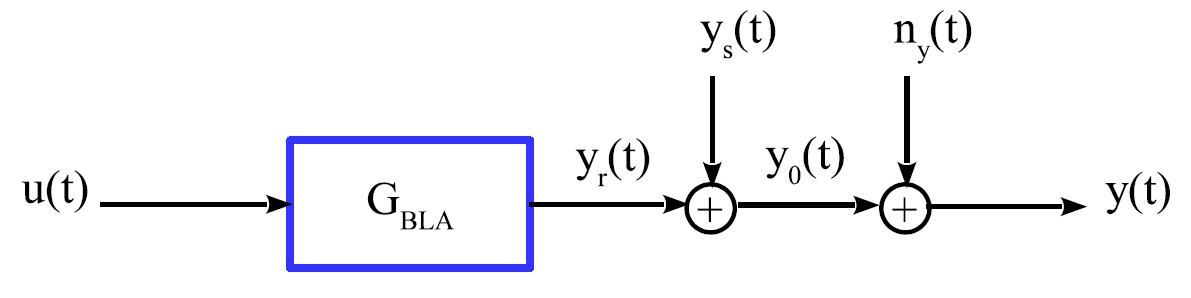}
	\caption{Measured input $u(t)$ and output $y(t)$ of a nonlinear system with the unmodeled nonlinear contributions $y_s(t)$ and the additive noise source $n_y(t)$.}
	\label{fig:system_bla}
\end{figure}

As mentioned earlier, the BLA framework offers a mathematical framework used in system identification to detect and quantify the nonlinear (NL) distortions in FRF measurements. In the BLA analysis, we consider the measurement setup as given in \refig{fig:system_bla}, where $u(t)$ and $y(t)$ are the input-output measurements. Below, we provide a brief introduction to the BLA framework. We refer the reader to \cite{schoukens} for a detailed introduction and analysis of the BLA framework.
\vspace{-0.5cm}

The BLA of a nonlinear system is given as
\begin{equation}
     G_{BLA}(q) \overset{\Delta}{=} \argmin_{G(q)} E\{|\tilde{y}(t)-G(q)\tilde{u}(t)|^2\},
\end{equation}
where $E\{.\}$ denotes the expected value operator \cite{schoukens}, and $q$ denotes the forward shift operator ($q u(t) = u(t+1)$). The zero-mean signals $\tilde{u}(t)$ and $\tilde{y}(t)$ are defined as
\begin{equation}
\begin{aligned}
     \tilde{u}(t) &= u(t) - E\{u(t)\}, \\
     \tilde{y}(t) &= y(t) - E\{y(t)\}. 
\end{aligned}
\end{equation}

The system labeled as \textit{Cell} in \refig{fig:system} is known to be a nonlinear dynamical system. Here, we characterize it by its best linear approximation $G_{BLA}$, as illustrated in \refig{fig:system_bla}:
\begin{equation}
y(t) = G_{BLA}(q) u(t) + y_s(t) + n_y(t),
\end{equation}
where the output equals the sum of the best linear approximation of the nonlinear system $G_{BLA}(q)$, distortion contributions due to noise ($n_y(t)$) and unmodeled nonlinearities ($y_s(t)$) existing within the system. Using the BLA framework, we can estimate the dynamics of $G_{BLA}$ and estimate the variances due to $y_s(t)$ and $n_y(t)$. 

The FRF of $G_{BLA}$ and the variances of distortion levels at each excited frequency can be estimated using the local polynomial method (LPM) \cite{schoukens, schoukens2018, schoukens2009}. LPM provides high-quality FRF estimates of a dynamical system that is excited by arbitrary or periodic input signals, starting from noisy measurements, while simultaneously estimating the transient/leakage contribution in the data which effectively improves the quality of the FRF estimate.

\section{Results}

\subsection{Measurements on Cells}

Using the measurement setup, experiments have been carried out on single cells employing multisine signals with different phase realizations. In the experiments, iPSC-derived vascular smooth muscle cells (iVSMCs) are used, sourced from a presumed healthy individual (after an informed consent procedure, as approved by the Maastricht University medical ethics committee). 
iVSMCs were maintained on 0.1$\%$ gelatin in DMEM-GlutaMAX (ThermoFisher, 10569010) at 37 degrees Celsius and 5$\%$ Co2. Cells were seeded sparsely in a 12-well plate, making sure single cells were separated sufficiently. For nanoindentation, the Pavone nanoindenter (Optics11Life) was preset at 37 degrees Celsius, 5$\%$ Co2, and 95$\%$ humidity by use of the environmental module add-on (Optics11Life). Measurements were performed by positioning the probe of the Pavone over the nucleus of the iVSMCs. 

Initially, a single measurement using a multisine signal was initiated on an individual cell, which denotes a sequence consisting of $500$ reference signal points, and $16000$ samples of the measured indentation and load due to the difference in sampling frequency during signal generation and measurement, equating to an approximate indentation period of $16$ seconds. Subsequently, we continued with our experiments by applying the identical signal on the same cell, in other words, we acquired multiple periods of measurements using the same signal. This approach is essential to characterize the noise level in the signals and to validate the variability of the measurements. Additionally, measurements employing different phase realizations of the multisine signal on the same cell are obtained for a thorough nonparametric characterization of the system's nonlinear behavior through the BLA framework. 

\subsection{Analysis of Measurements}

\begin{figure}[h!]
	\centering
	\includegraphics[width=0.82\linewidth]{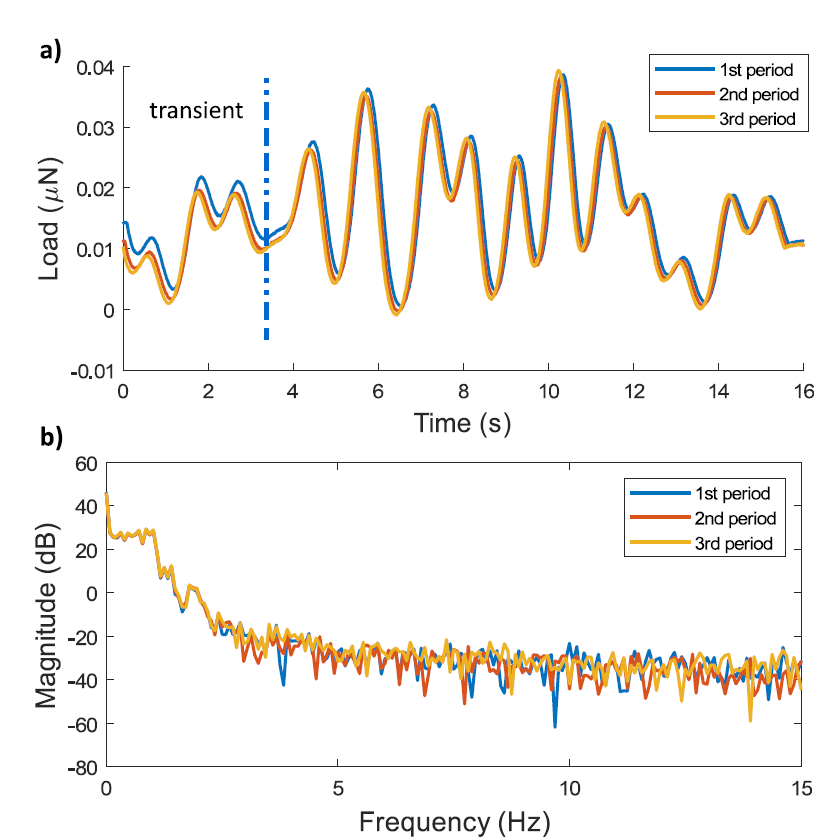}
	\caption{a) 3 measured periods of load data for one phase realization of the multisine reference signal. b) Load data DFT spectrum for each of the 3 periods.}
	\label{fig:load1}
\end{figure}

\begin{figure}[h!]
	\centering
	\includegraphics[width=0.82\linewidth]{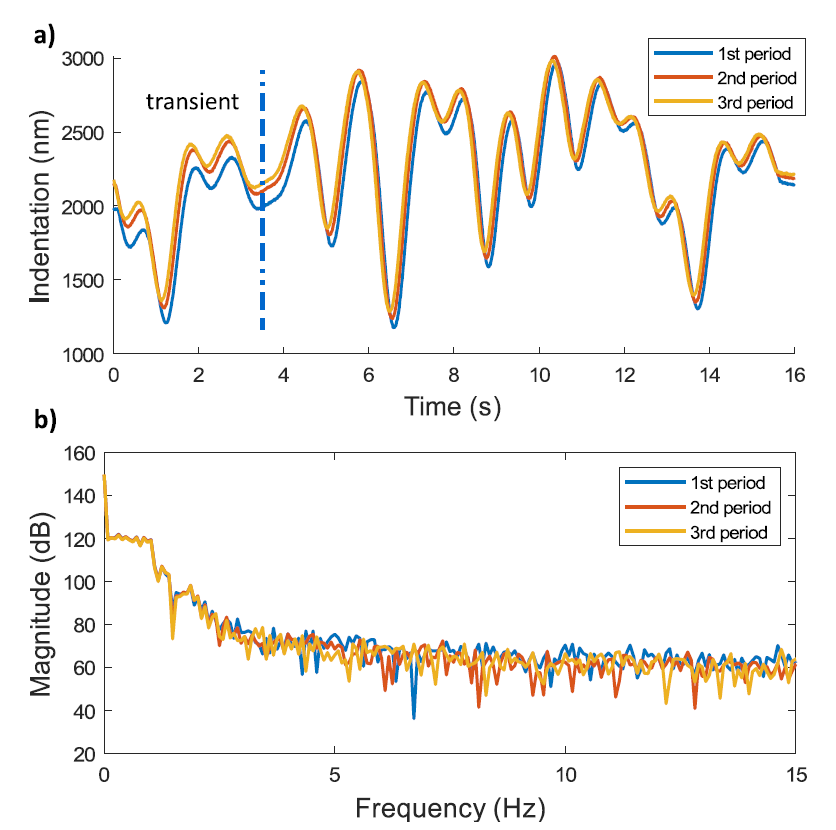}
	\caption{a) 3 measured periods of indentation data for one phase realization of the multisine reference signal. b)Indentation data DFT spectrum for each of the 3 periods.}
	\label{fig:out1}
\end{figure}

A single iVSMC is subjected to a multisine signal characterized by a linear frequency distribution in the range $[f_{min}, f_{max}] = [0.06 \, \text{Hz}, 1 \, \text{Hz}]$, using a sampling frequency of $f_s = 31.25 $ Hz.  
Three periods of load signal measurements ($u(t)$) are shown \refig{fig:load1} (a), together with their DFT spectra in \refig{fig:load1} (b). Similarly, the $3$ periods of indentation measurements ($y(t)$) and their corresponding DFT spectra are shown in \refig{fig:out1} (a) and \refig{fig:out1} (b) respectively. 
The measurements for both load and indentation appear periodic, after the ~2s transient time, validating the experiment design choices outlined in the previous sections of this paper. 
It should be mentioned that the results shown here represent the raw data, collected directly through the measurement setup, no preprocessing of the data has taken place.

\begin{figure}[t!]
	\centering
	\includegraphics[width=0.9\linewidth]{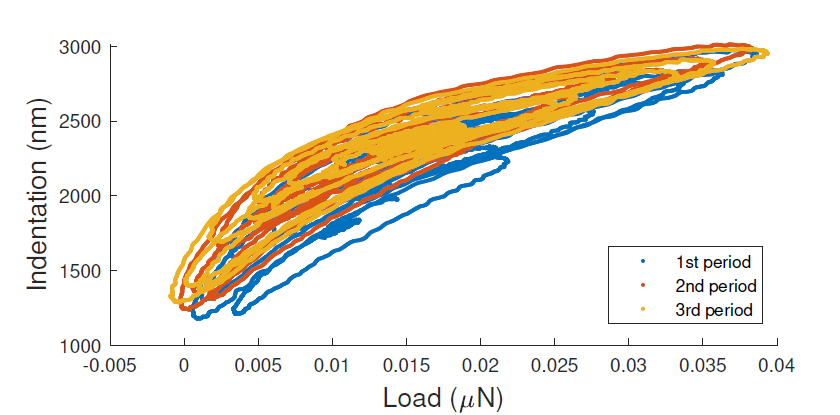}
	\caption{Load and indentation data of a healthy cell for 3 periods of the multisine.}
	\label{fig:iVSMC001_cell1_load_indentation_3periods}
\end{figure}

\begin{figure}[t!]
	\centering
	\includegraphics[width=0.9\linewidth]{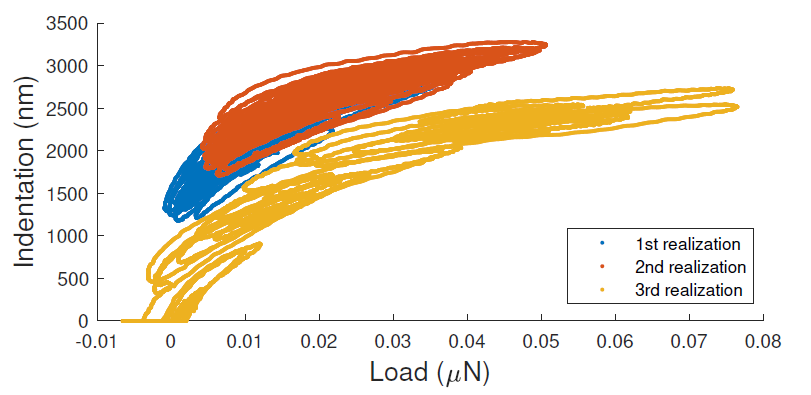}
	\caption{3 periods of the load and indentation data of a healthy cell with different realizations.}
\label{fig:iVSMC001_cell1_load_indentation_3realizations}
\end{figure}

\refig{fig:iVSMC001_cell1_load_indentation_3periods} shows the load versus indentation measurements of different periods, which exhibit minimal degradation and demonstrate a resemblance to one another. There appears to be a potential nonlinear relation between load and indentation, together with some dynamics.  
Following this, our analysis continued by employing various phase realizations of the multisine signal on the same single cell to characterize this nonlinear behavior. We then observed notable degradation over realizations.
During the application of the third multisine realization, the cell's response begins to deteriorate as seen in \refig{fig:iVSMC001_cell1_load_indentation_3realizations}. 
This degradation emphasizes the need for fast measurement procedures, such as the use of a multisine signal.
Furthermore, note that the measurements from the initial two realizations align closely on the same trajectory, although there is a small shift in the set point, indicating a shift in the load data. 
Due to the observed degradation, only the measurements obtained using the first two realizations of the multisine signal are utilized in the subsequent frequency-domain analysis. 

\begin{figure}[b]
	\centering
	\includegraphics[width=\linewidth]{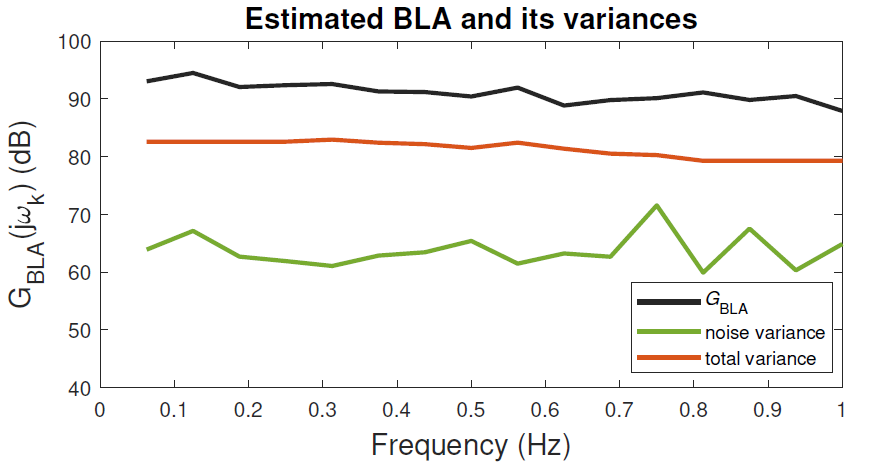}
	\caption{Measurement of the best linear approximation with a healthy cell. Black: Amplitude of the BLA. Green: Variance of the BLA estimate due to noise. Red: The total variance (noise+nonlinearities) of the BLA estimate.}
	\label{fig:iVSMC001_cell1_bla}
\end{figure}

We have analyzed the BLA estimate of the system over a frequency range $[0.06 \, \text{Hz}, 1 \, \text{Hz}]$. The LPM utilizes $P = 3$ consecutive periods of the data and $M = 2$ phase realizations. The results indicate a low level of noise distortions has been observed along with significant nonlinearities. As can be seen in \refig{fig:iVSMC001_cell1_bla} the variance due to noise is approximately $30$ dB below the FRF estimate while the SNR due to the total variance, including both noise and nonlinear distortions, is only $10$ dB below the FRF estimate. When projected back to the output, this indicates that nearly $30 \%$ of the system response is due to nonlinearities for these types of measurements. Note that the nonlinearity level is estimated using only $2$ realizations of the multisine signal in the analysis with a single cell. Hence, the estimated nonlinearity level contains some uncertainty. Further measurements will be required to obtain a more accurate characterization.   

\subsection{Discussions}

We have focused on the analysis of SMC dynamics at the level of iPSC-derived healthy cells using a random phase multisine signal with a linear frequency distribution.    
Load and indentation measurements are examined over various periods and different phase realizations using an individual cell, as illustrated in \refig{fig:iVSMC001_cell1_load_indentation_3periods} and \refig{fig:iVSMC001_cell1_load_indentation_3realizations}. Minimal degradation is observed between the measurements across different periods. However, a shift is noted in the load data as a result of conducting repeated measurements with different phase realizations. 

Some dynamic behavior is observed in the analysis, with these dynamics being predominantly smooth. Correspondingly, noticeable nonlinearities are evident in the frequency spectrum analysis based on the results obtained with the first two phase realizations of the multisine. The BLA analysis shown in \refig{fig:iVSMC001_cell1_bla} confirms the presence of dynamics as a slight slope in the amplitude of $G_{BLA}$ is observed. We consider these dynamics to reflect (i) recruitment/alignment of cytoskeletal fibers (curvi-linearity), and (ii) if not due to friction, due to fluid shifts in the cell body, possibly a myogenic modulation of the load-indentation relationship.   

\section{Conclusion}

In this paper, we have introduced a measurement setup and an identification procedure for analyzing the frequency-domain dynamics of vascular SMCs. The main goal of this study was to develop the measurement procedure and to demonstrate the feasibility of using a nano-indentation measurement setup for the broadband frequency-domain characterization of the mechanical properties of SMCs. The results indicated that the frequency-domain dynamics of SMCs are characterized with high precision, and the nonlinear behavior is quantified, using random phase multisine signals in combination with the local polynomial FRF estimation method.  

Our future outlook is, by using this identification procedure, to perform measurements on a population of cells to characterize the dynamics and the variability of measurements both within a single cell and across multiple cells. Moreover, we seek to characterize and compare the dynamics of paired disease and naive SMCs obtained from aortic aneurysm patients to understand diseased tissue developed at the cellular level.

\end{document}